\documentclass[twocolumn,twoside,slac_two]{revtex4}
\usepackage{graphicx}
\usepackage{fancyhdr}

\newcommand{\HI}{{\sc H}\,{\scriptsize{\sc I}}}
\newcommand{\Av}{$A{\rm v}$}
\newcommand{\Avres}{$A{\rm v_{res}}$}
\newcommand{\Wco}{$W_{\rm CO}$}

\newcommand{\Xco}{$X_{\rm CO}$}
\newcommand{\XAvres}{$X_{\rm Avres}$}
\newcommand{\bi}{\bfseries\itshape}

\begin{document}

\title{{\bi Fermi} LAT Study of the Cosmic-rays and the Interstellar Medium in Nearby Molecular Clouds}

%

\author{K. Hayhashi$^{a}$ and T. Mizuno$^{b}$ on behalf of the 
{\it Fermi}-LAT collaboration}
\affiliation{$^{a}$Department of Physical Science, Hiroshima University, 
Higashi-Hiroshima, Hiroshima 739-8526, Japan \\
$^{b}$Hiroshima Astrophysical Science Center, Hiroshima University, 
Higashi-Hiroshima, Hiroshima 739-8526, Japan}
%

\begin{abstract}
We report an analysis of the interstellar gamma-ray emission from 
nearby molecular clouds Chamaeleon, R~Coronae Australis (R~CrA), 
and Cepheus and Polaris flare regions with the {\it Fermi} Large Area 
Telescope (LAT). They are
among the nearest molecular cloud complexes, within $\sim$ 300 pc from the 
solar system. The gamma-ray emission produced by interactions of 
cosmic-rays (CRs) and interstellar gas in those molecular clouds is useful 
to study the CR densities and distributions of molecular gas
close to the solar system.
The obtained gamma-ray emissivities from 250 MeV to 
10 GeV for the three regions are 
about (6--10)  
$\times$ 10$^{-27}$ photons s$^{-1}$ sr$^{-1}$ H-atom$^{-1}$,
indicating a variation of the CR density by $\sim$ 20 \% 
even if we consider the systematic 
uncertainties. 
The molecular mass calibration ratio, 
$X_{\rm CO} = N{\rm (H_2)}/W_{\rm CO}$, is found to be 
about (0.6--1.0) $\times$ 10$^{20}$ H$_2$-molecule cm$^{-2}$       
(K km s$^{-1}$)$^{-1}$ among the three regions, 
suggesting a variation of $X_{\rm CO}$ in the vicinity 
of the solar system. From the obtained values of $X_{\rm CO}$, 
we calculated masses of molecular gas traced by \Wco\ in these molecular
clouds. In addition, similar amounts of dark gas at the interface between
the atomic and molecular gas are inferred.
\end{abstract}

\maketitle

\thispagestyle{fancy}

\section{INTRODUCTION}
Observation of high-energy gamma-rays from molecular clouds can be used to 
study the sources of
cosmic rays (CRs), the CR density, and the distribution of the
interstellar medium (ISM) in such systems. Gamma-rays are produced in
the ISM by interactions of high-energy CR protons and electrons with
the interstellar gas or the interstellar radiation field, via
nucleon-nucleon collisions, electron bremsstrahlung, and inverse
Compton (IC) scattering. Since the gamma-ray production cross section
is almost independent of the thermodynamic state of the ISM,
gamma-rays have been recognized as a powerful probe of the
distribution of interstellar matter (e.g., Stecker 1970,
Lebrun et al. 1983 and Blemen et al. 1984)
If the gas
column densities are estimated with good accuracy by observations in
other wavebands such as radio and infrared, the CR spectrum and
density distributions can be examined as well.
Molecular clouds that are within 1 kpc from the solar system 
(namely nearby molecular clouds) and
have masses greater than a few 10$^3$ {\it M}$_{\odot}$
are well suited for an analysis of their
gamma-ray emission to investigate the distribution of CR densities and
interstellar gas since they are observed at high latitudes and
therefore largely free from confusion with the strong emission from
the Galactic plane. 
Study of such nearby molecular clouds in gamma-rays can be dated back to
the COS-B era (Bloemen et al. 1984) 
and was advanced by the EGRET
on board {\it Compton Gamma-Ray Observatory} 
(e.g., Hunter et al. 1994).
Although some important information has been obtained on
properties of CRs and the ISM by these early observations, detailed
studies have only been performed on giant molecular clouds with masses
greater than $\sim$~10$^5$ {\it M}$_{\odot}$ such as the Orion complex
(e.g., Digel et al. 1999).
The Large Area Telescope (LAT) 
(Atwood et al. 2009)
on board the {\it Fermi 
Gamma-ray Space Telescope}
has improved the situation significantly thanks to its unprecedented
sensitivity. In addition, recent developments in studies of the ISM
(e.g., Kalberla et al. 2005, Dame et al. 2001 and Grenier et al. 2005)
allow us to investigate the CR spectra and density
distribution with better accuracy.

Here, we report an analysis of {\it Fermi} LAT observations of the
Chamaeleon, R~Coronae Australis (R~CrA), and Cepheus and Polaris flare
molecular clouds. They are among the nearest ($\sim$ 300 pc from the solar
system) molecular clouds. No detailed study of CR and matter
distributions for the Chamaeleon and R~CrA regions had been performed
before since they have rather small masses ($\lesssim$ 10$^4$ $M_{\odot}$) 
and consequently small gamma-ray fluxes. We also analyzed in detail 
the region of the
Cepheus and Polaris flares which was included in the {\it Fermi} LAT study
of the second Galactic quadrant 
(Abdo et al. 2010a).
Details of this proceedings are presented in a published paper 
(Ackermann et al. 2012a).

\section{DATA ANALYSIS}
\begin{figure*}[t]
\begin{tabular}{ll}
\begin{minipage}{0.5\hsize}
\includegraphics[width=80mm]{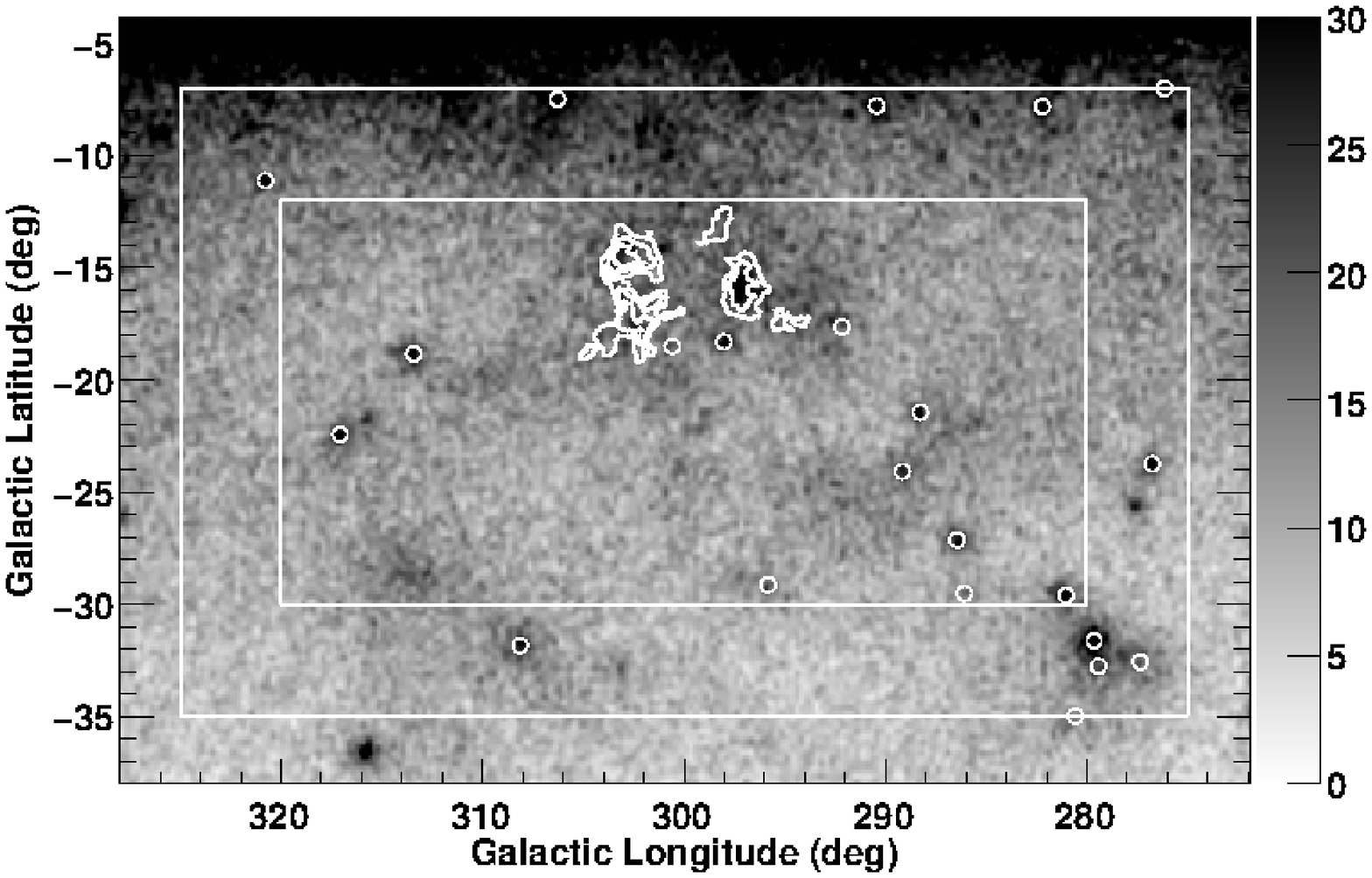}
\end{minipage}
\begin{minipage}{0.5\hsize}
\includegraphics[width=80mm]{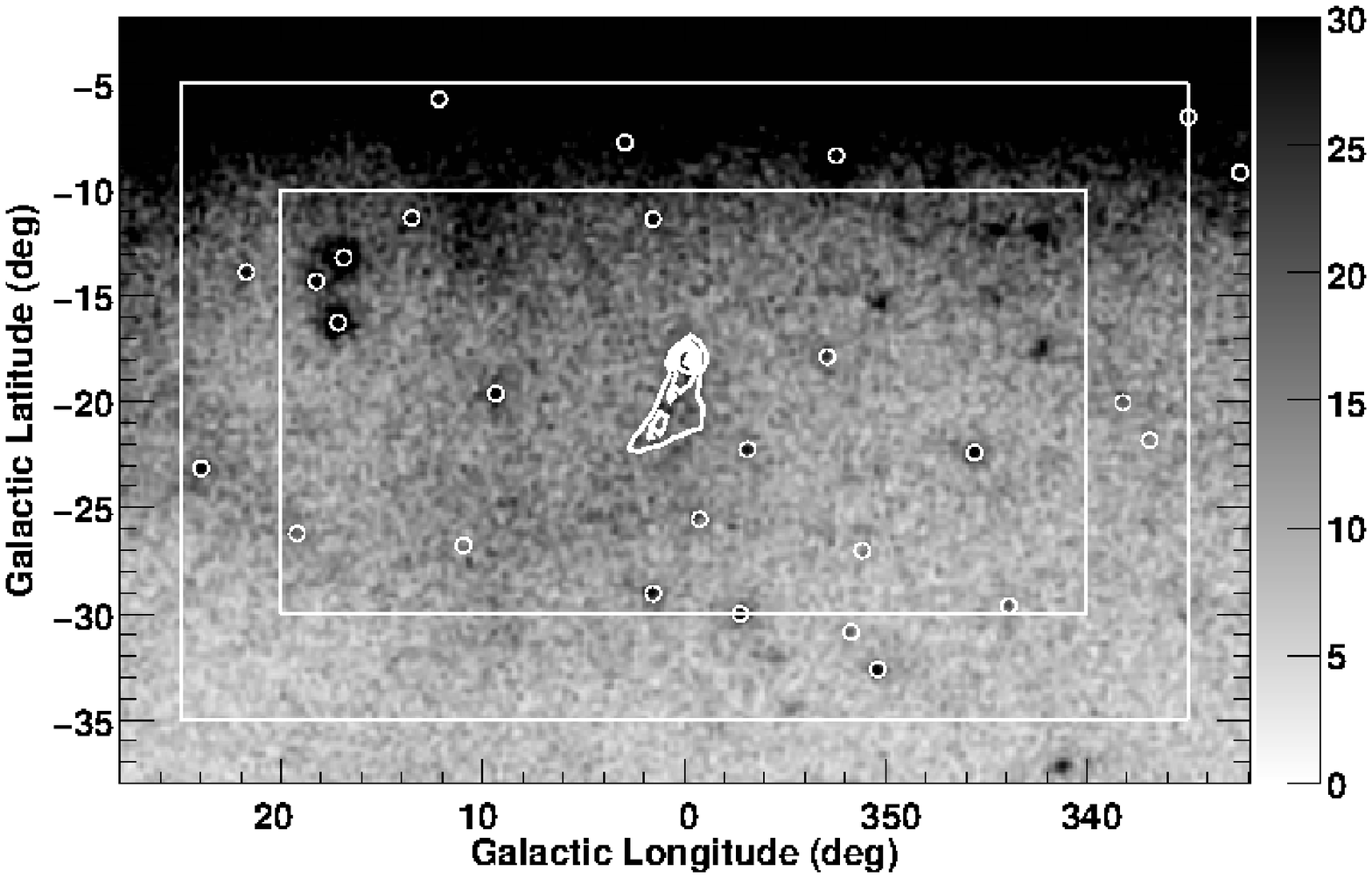}
\end{minipage}\\
\begin{minipage}{0.5\hsize}
\includegraphics[width=80mm]{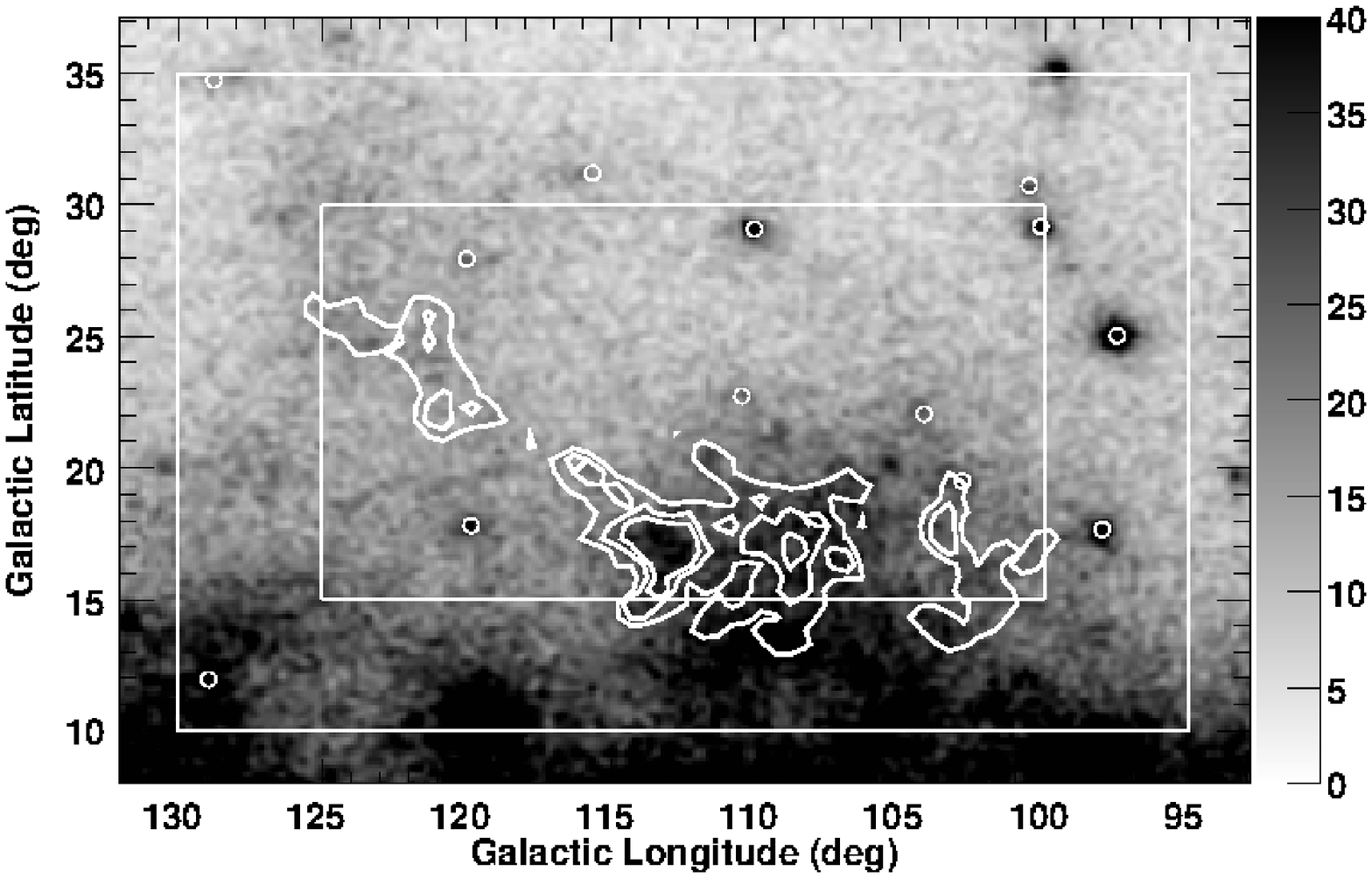}
\end{minipage}
\begin{minipage}{0.5\hsize}
\end{minipage}\\
\end{tabular}
\caption{Gamma-ray count maps above 250 MeV for the Chamaeleon (top
  left), R~CrA (top right), and Cepheus and Polaris flare (bottom
  left) regions, smoothed with a Gaussian of $\sigma =$ 0.5$^{\circ}$
  for display. The contours indicate intensities $W_{\rm CO}$ of the
  2.6 mm line of CO (with the levels of 4, 8, 12, and 16 K km
  s$^{-1}$) by Dame et al. (2001), 
  as a standard tracer of the molecular 
  gas. White circles show the positions of point sources with high
  significance (TS $\geq$ 50) in the First {\it Fermi} LAT catalog
  (1FGL) by Abdo et al. (2010b). 
  The inner squares indicate the ROI
  analyzed for each of the regions. Point sources outside of this ROI
  but inside the outer square are taken into account in the analysis.}
\label{fig:Cham_R_CrA_CePo_image}
\end{figure*}

For this analysis, we have accumulated events obtained from 2008 August 4 
to 2010 May 9. During this time interval the LAT was operated in sky survey 
mode nearly all of the time and scanned the gamma-ray sky with relatively
uniform exposure over time (within 10\% in regions studied).
We used the standard LAT analysis software,
Science Tools\footnote{Available from the {\it Fermi} Science Support Center
(http://fermi.gsfc.nasa.gov/ssc/).} version v9r16p0 and the response
function P6\_V3\_DIFFUSE, which was developed to account for the
detection inefficiencies due to pile-up and accidental coincidence of
events (Rando et al. 2009).
We set the lower energy limit at 250 MeV
to utilize good angular resolution (68\% containment
radius is $\lesssim$ 1.5$^{\circ}$ above 250 MeV)
and the upper energy limit at 10 GeV because of limited photon statistics.
The gamma-ray count maps obtained ($E$ $>$ 250 MeV) in the three regions are
shown in Figure \ref{fig:Cham_R_CrA_CePo_image}.
Point sources with high significance (test
statistic, TS\footnote{TS is defined as
${\rm TS} = 2({\rm ln}L-{\rm ln}L_{0})$, where $L$ and $L_{0}$ are the
maximum likelihoods obtained with and without the source included in the model
fitting, respectively; see (Mattox et al. 1996)},
greater than 50) are also plotted.

We modeled the gamma-ray emission as a linear combination of maps
tracing the column density of the interstellar gas. This approach has
been successfully applied in recent studies of diffuse gamma-rays by
the LAT (e.g., Abdo et al. 2010a and Ackermann et al. 2011). 
With the usual assumptions of optical
thinness and that CRs uniformly thread the ISM, gamma-ray intensity
(s$^{-1}$ cm$^{-2}$ sr$^{-1}$ MeV$^{-1}$) at a given energy can be modeled as


\begin{eqnarray*}
I_{\gamma}(l,b) & = & \sum_{i=1}^{2} q_{{\rm H I},i} 
\cdot N({\rm H_{\ I}})(l,b)_{i} + q_{\rm CO} \cdot W_{\rm CO}(l,b) \\
                & + & q_{\rm Av} \cdot A{\rm v}_{\rm res}(l,b) 
                + {\rm c_{IC}}\cdot I_{\rm IC}(l,b) \\
                & + & I_{\rm iso} + \sum_{j} {\rm PS}_{j} 
\label{eq:fit_model}
\end{eqnarray*}
where $i$ labels the regions along the line of sight separated in the
analysis (1 for the Gould Belt and 2 for the rest). $q_{{\rm HI},i}$ 
(s$^{-1}$ sr$^{-1}$ MeV$^{-1}$) and $q_{\rm CO}$ (s$^{-1}$ cm$^{-2}$ 
sr$^{-1}$ MeV$^{-1}$ (K km s$^{-1}$)$^{-1}$) are the emissivity
per \HI\ atom (traced by the 21 cm line of atomic hydrogen) and per 
$W_{\rm CO}$ unit (as a tracer of molecular gas), respectively. 
$q_{\rm Avres}$ (s$^{-1}$ cm$^{-2}$ sr$^{-1}$ MeV$^{-1}$ mag$^{-1}$) 
is the emissivity per \Avres\ magnitude which accounts
for the gas not traced well by \HI\ and CO: we constructed visual
extinction (\Av) maps providing an estimate of the
total column densities on the assumption of a constant gas-to-dust
ratio, and obtained  map by the fitting a linear combination of
the $N({\rm HI})$ and \Wco\ maps. $I_{\rm IC}$ and $I_{\rm iso}$ are
the IC model and isotropic
background intensities (s$^{-1}$ cm$^{-2}$ sr$^{-1}$ MeV$^{-1}$), 
respectively. We used
GALPROP\footnote{http://www.mpe.mpg.de/\~{}aws/propagate.html} 
(e.g., Strong and Moskalenko 1998) for the IC model,
and a publicly available
isotropic spectrum\footnote{isotropic\_iem\_v02.txt from
 http:\slash\slash{}fermi.gsfc.nasa.gov\slash{}ssc\slash{}data\slash{}access\slash{}lat\slash{}BackgroundModels.html} 
for the isotropic component. PS$_{j}$ represents
contributions of point sources included in the 1FGL catalog
(Abdo et al. 2010b).
See Ackermann et al. (2012a) for details of the analysis. 

\section{RESULTS}

\begin{figure*}[t]
\begin{tabular}{cc}
\begin{minipage}{0.5\hsize}
\includegraphics[width=80mm]{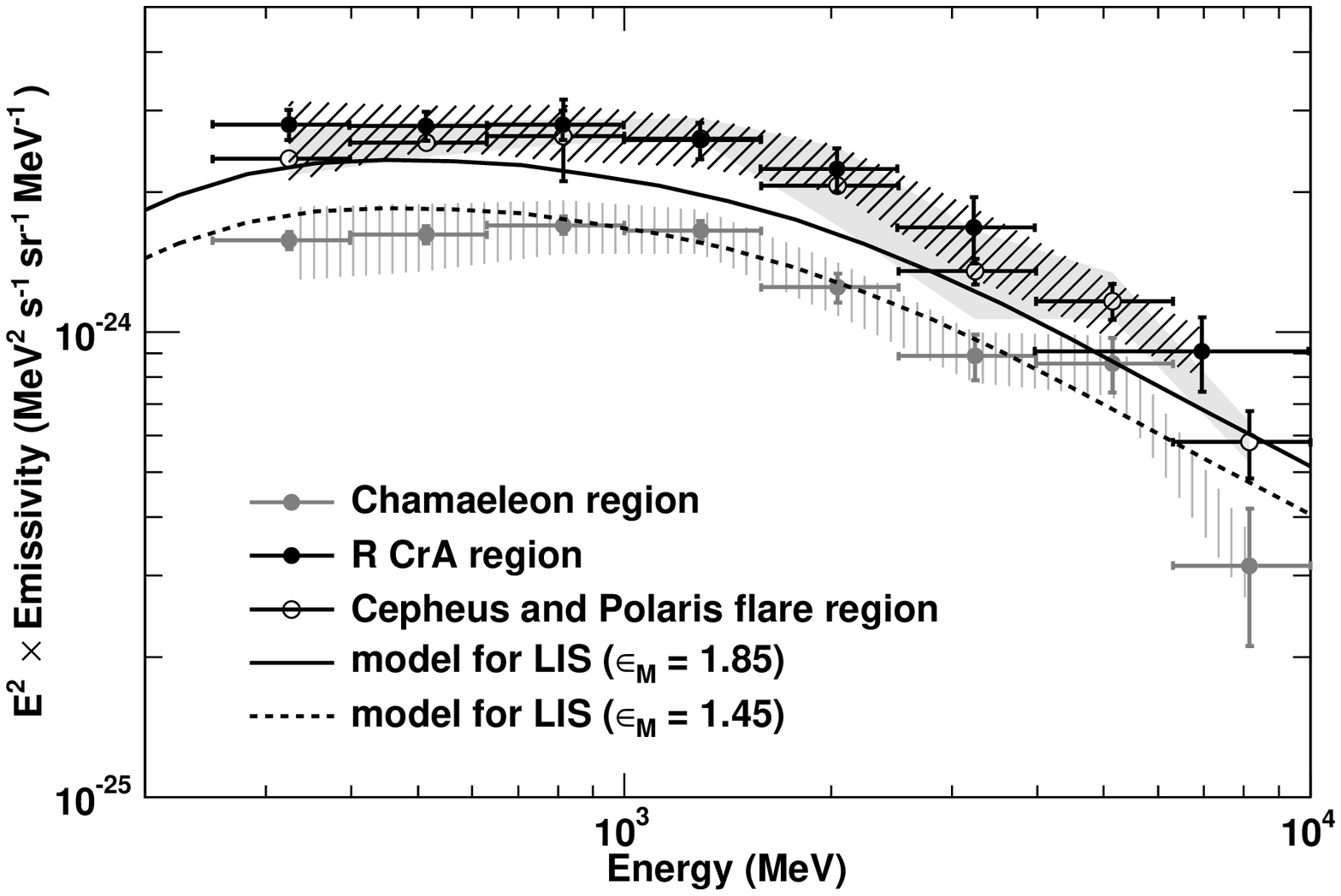}
\end{minipage}
\begin{minipage}{0.5\hsize}
\includegraphics[width=80mm]{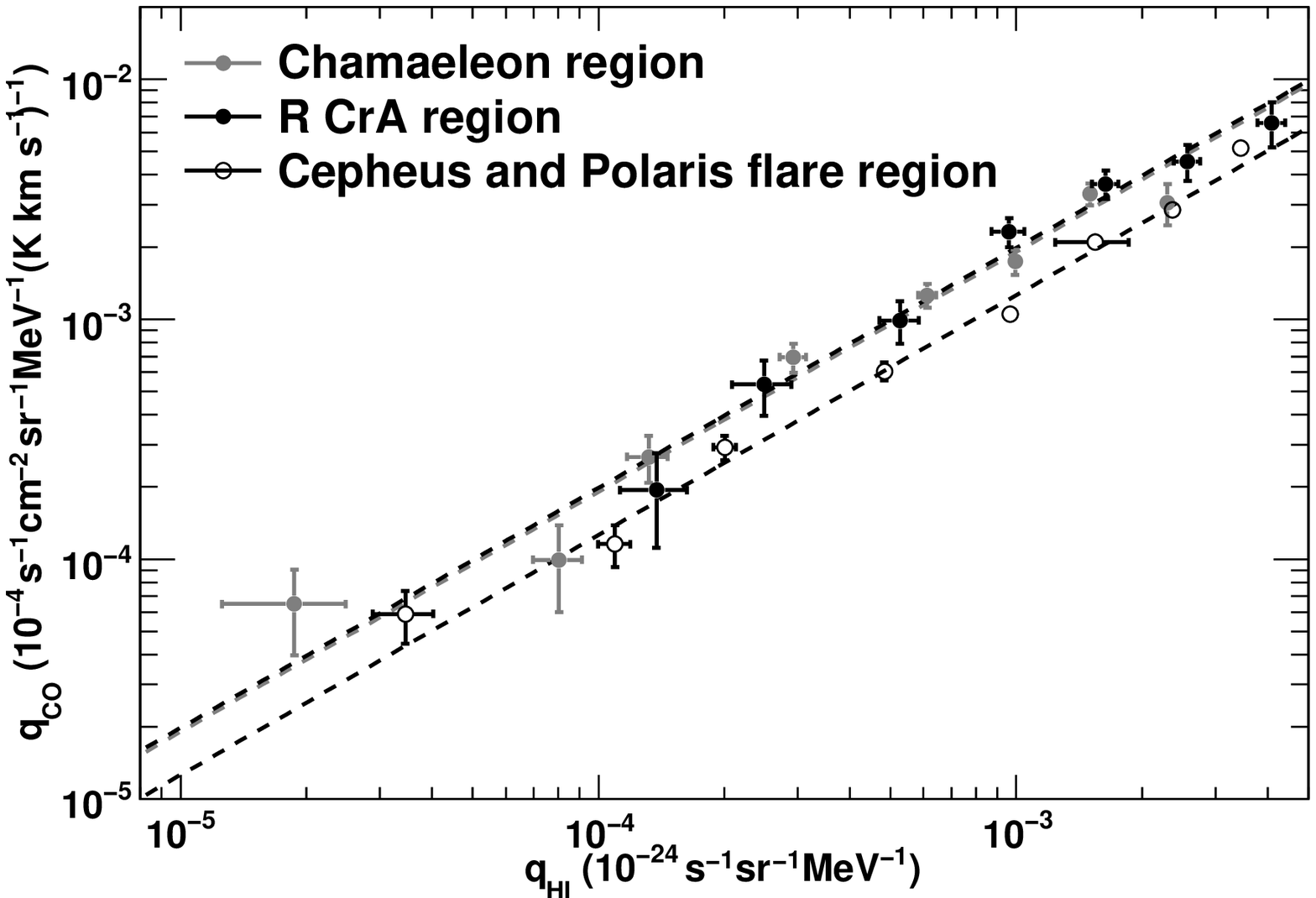}
\end{minipage}\\
\end{tabular}
\caption{(left): \HI\ emissivity spectra of each region compared with
  the model for LIS with the nuclear enhancement factors of 1.45 and
  1.84 (Mori 2009). 
  The shaded areas indicate the systematic
  uncertainty: vertical lines (Chamaeleon), diagonal lines (R~CrA) and 
  the gray area (Cepheus and Polaris flare).  
  (right): CO versus \HI\ emissivities. Each point corresponds to an 
  energy bin. Errors are statistical only.}
\label{fig:HIemissivity_Xco}
\end{figure*}

Figure \ref{fig:HIemissivity_Xco} (left) shows the obtained gamma-ray 
emissivity spectra per
H-atom of each region assuming the spin temperature $T_{\rm s}$ $=$ 125 K. The
emissivities above 250 MeV are 
(5.9 $\pm$ 0.1$_{\rm stat}$ $^{+0.9}_{-1.0}$$_{\rm sys}$) 
$\times$ 10$^{-27}$ photons s$^{-1}$ sr$^{-1}$ H-atom$^{-1}$, 
(10.2 $\pm$ 0.4$_{\rm stat}$ $^{+1.2}_{-1.7}$$_{\rm sys}$) $\times$ 10$^{-27}$ photons 
s$^{-1}$ sr$^{-1}$ H-atom$^{-1}$, and
(9.1 $\pm$ 0.3$_{\rm stat}$ $^{+1.5}_{-0.6}$$_{\rm sys}$) $\times$
10$^{-27}$ photons s$^{-1}$ sr$^{-1}$ H-atom$^{-1}$
for the Chamaeleon, R~CrA, and Cepheus
and Polaris flare regions, respectively. 
In order to examine the systematic uncertainty due to the optical depth
correction, we also tried to fit the data with maps obtained by assuming
{\it T}${\rm _s}$ = 100 K and under the approximation that the gas is
optically-thin. We evaluated the uncertainty of the isotropic component
to be $\pm$10\%
by comparing the model we adopted and those derived in
other LAT studies of mid-latitude regions 
(Abdo et al. 2009a and Abdo et al. 2009b).
We thus reran the analysis 
assuming a 10\% higher and lower intensity for the fixed
isotropic component.
We also investigated the effect on the systematic uncertainty due to the
IC component by using different IC model maps which are constructed under
different assumptions about the distribution of CR sources and intensity
of the interstellar radiation field 
(e.g., Ackermann et al. 2012b).
The effects of the
uncertainty of the {\it T}${\rm _s}$, isotropic component, and IC models
are quite comparable, therefore we added them.
The resultant peak-to-peak uncertainty of the local {\HI} emissivity is less
than $\sim$ 20 \% across the energy range for three regions investigated.

\section{DISCUSSION}
\subsection{CR Density and Spectrum Close to the Solar System}

We then compared the obtained \HI\ emissivity spectra among the three
regions, and model spectra\footnote{The model is calculated from the LIS
compatible with the CR proton spectrum measured by 
Alcaraz et al. (2000) and Sanuki et al. (2000),
under the assumption that the nuclear
enhancement factors are 1.45 and 1.84 
(Mori 2009); see Abdo et al. (2009b).} 
for the local interstellar spectrum (LIS) as shown in Figure 
\ref{fig:HIemissivity_Xco} (left).
Whereas the spectral shapes for the three regions
studied here agree well with the LIS model, the absolute emissivities
differ among the three regions. 
We note that the systematic uncertainty of the LAT effective area
(5\% at 100 MeV and 20\% at 10 GeV; 
(Rando et al. 2009)
does not affect the relative value of emissivities
among these regions. 
The effect of unresolved point sources is small, since we
have verified that the obtained emissivities are robust against the
lower threshold for point sources between TS $=$ 50 and 100.
Although the emissivities of the R~CrA
region and the Cepheus and Polaris flare region are comparable, that
of the Chamaeleon region is lower by $\sim$ 20\%, even if we take the
systematic uncertainties into account. Therefore, the LAT data point
to variations in the CR densities within $\sim$ 300 pc in the Gould Belt. 
If the CR density has a variation by a factor of $\sim$ 1.2 in the
neighborhood of the solar system, this requires a serious reconsideration
of a smooth CR density often
adopted for simplicity, and may have an impact on the study of the CR
source distribution and diffuse gamma-ray emission.
We note that CR sources are stochastically distributed in space and time,
and this may produce a CR anisotropy depending on the propagation
conditions (e.g., Blasi \& Amado 2012a and Blasi \& Amato 2012b).
Study of other regions and more detailed
theoretical calculations are needed for further investigating this
issue. On the other hand, we need to further investigate possible
systematic uncertainties due to variations in \Xco\ (see the next
paragraph) or dust-to-gas ratio we assumed in the construction of the
\Avres\ map.

\subsection{Molecular Masses in the Interstellar Clouds Studied}

Under the hypothesis that the same CR flux penetrates the \HI\ and CO
phases of an interstellar complex, we can calculate the molecular mass
calibration ratio, \Xco, as 
$X_{\rm CO} = q_{\rm CO}/(2q_{\rm HI})$. As shown in 
Figure \ref{fig:HIemissivity_Xco} (right), we derived the \Xco\ values from a
linear fit of the ($q_{\rm HI}$, $q_{\rm CO}$) points. 
The obtained \Xco\ values are summarized in Table \ref{table:mass}:
(0.96 $\pm$ 0.06$_{\rm stat}$ $^{+0.15}_{-0.12}$$_{\rm sys}$) 
$\times$ 10$^{20}$ cm$^{-2}$ (K km s$^{-1}$)$^{-1}$,
(0.99 $\pm$ 0.08$_{\rm stat}$ $^{+0.18}_{-0.10}$$_{\rm sys}$) 
$\times$ 10$^{20}$ cm$^{-2}$ (K km s$^{-1}$)$^{-1}$, and
(0.63 $\pm$ 0.02$_{\rm stat}$ $^{+0.09}_{-0.07}$$_{\rm sys}$) 
$\times$ 10$^{20}$ cm$^{-2}$ (K km s$^{-1}$)$^{-1}$
for the Chamaeleon, R~CrA, and Cepheus and Polaris flare regions, respectively.
The obtained value of \Xco\ for the Cepheus and Polaris flare region is
$\sim$ 20 \% lower than that reported by Abdo et al. (2010a). 
Abdo et al. (2010a) includes in their study also the Cassiopeia
molecular cloud in the Gould Belt, and due to different ROIs considered,
the $q_{\rm HI}$ emissivity was also different.
\Xco\ of the Chamaeleon region is similar to that of the R~CrA region,
whereas that of the Cepheus and Polaris flare region is $\sim$2/3 of the
others. The LAT data thus suggest a variation of \Xco\
on a $\sim$ 300 pc scale.

Using these \Xco\ values, the molecular masses traced by
\Wco\ can be calculated. The mass of the gas traced by \Wco\ is expressed as
\begin{eqnarray*}
\frac{M}{M_{\odot}} = 2 \mu\frac{m_{\rm H}}{M_{\odot}} d^{2} X_{\rm CO} \int
 W_{\rm CO}(l,b)\ d\Omega
\label{eq:mass}
\end{eqnarray*}
where $d$ is the
 distance to the cloud, $m_{\rm H}$ is the mass of the
hydrogen atom and $\mu = 1.36$ is the mean atomic weight per H-atom
(Allen 1973).
From this equation the mass of gas traced by CO
is expressed as $M_{\rm CO}$ in Table \ref{table:mass}:
we obtained $\sim$ 5$\times$10$^{3}$ $M_{\odot}$, $\sim$ 
10$^{3}$ $M_{\odot}$, and
$\sim$ 3.3$\times$10$^{4}$ $M_{\odot}$
for the Chamaeleon, R~CrA, and Cepheus and Polaris flare regions,
respectively.
From the linear relation between $q_{\rm HI}$ and $q_{\rm Avres}$ 
($q_{\rm Avres} = X_{\rm Avres} \cdot q_{\rm HI}$) we
can calculate the masses of additional gas traced by $q_{\rm Avres}$ 
with a procedure similar to that for CO and results are summarized in 
Table~\ref{table:mass}. 
We thus obtained mass estimates for the Chamaeleon and R~CrA regions 
similar to previous ones (Dame et al. 1987) 
if we consider the total mass (traced by \Wco\ and
\Avres),
valthough the procedure is not straightforward since the gas traced by
\Avres\ is extended in a much larger region of the
sky. Detailed study of the matter distribution in the interstellar space
by comparing gamma-rays and other tracers will be reported elsewhere.

\begin{table*}[ht]
 \caption{\Xco, \XAvres\ and masses in the interstellar clouds in each region}
  \begin{center}
  \label{table:mass}
   \begin{tabular}{ccccc}\hline\hline
        & \Xco  & $M_{\rm CO}$  & \XAvres & $M_{\rm Avres}$ \\
        & ($\times$ 10$^{20}$ cm$^{-2}$ (K km s$^{-1}$)$^{-1}$) 
                & ($M_{\odot}$) & ($\times$ 10$^{22}$ cm$^{-2}$ mag$^{-1}$)
                & ($M_{\odot}$) \\ \hline
   Chamaeleon & 0.96$\pm$0.06$_{\rm stat}$$\pm$0.13$_{\rm sys}$ &
                $\sim$5$\times$10$^3$ &
                0.22$\pm$0.01$_{\rm stat}$$\pm$0.08$_{\rm sys}$ &
                $\sim$2.0$\times$10$^4$ \\
   R~CrA      & 0.99$\pm$0.08$_{\rm stat}$$\pm$0.14$_{\rm sys}$ &
                $\sim$10$^3$ &
                0.21$\pm$0.01$_{\rm stat}$$\pm$0.02$_{\rm sys}$ &
                $\sim$10$^3$ \\
   Cepheus \& Polaris & 0.63$\pm$0.02$_{\rm stat}$$\pm$0.08$_{\rm sys}$ &
                $\sim$3.3$\times$10$^4$ &
                0.14$\pm$0.01$_{\rm stat}$$\pm$0.03$_{\rm sys}$ &
                $\sim$1.3$\times$10$^4$ \\
   \hline 
   \end{tabular}
  \end{center}
\end{table*}

\section{Summary and Conclusions}
\label{sec:Summary_and_conclusions}

We have studied the gamma-ray emission from the Chamaeleon, R~CrA,
and Cepheus and Polaris flare molecular clouds close to the solar 
system ($\lesssim$ 300 pc) using
the first 21 months of {\it Fermi} LAT data. Thanks to the excellent
performance of the LAT, we have obtained unprecedentedly high-quality
emissivity spectra of the atomic and molecular gas in these regions in
the 250 MeV -- 10 GeV range.

The gamma-ray emissivity spectral shapes in three regions
agree well with the model for the LIS (a model based on local CR measurement),
thus indicating a similar spectral distribution of CRs in these regions.
The emissivities, however, indicate a variation of the CR
density of $\sim$ 20~\% within $\sim$ 300 pc around the solar system,
even if we consider the systematic uncertainties.
We consider possible origins of the variation are non-uniform
supernova rate and anisotropy of CRs depending on the propagation conditions.

The molecular mass calibration ratio \Xco\ for
the Chamaeleon cloud and the
R~CrA cloud are comparable, whereas that of the Cepheus and Polaris flare
region is $\sim$ 2/3 of the others,
suggesting a variation of
\Xco\ in the vicinity of the solar system. From the obtained
values of \Xco, the masses of gas traced by
\Wco\ in the Chamaeleon, R~CrA, and Cepheus and Polaris flare regions
are estimated to be $\sim$ 5 $\times$ $10^{3}$ $M_{\odot}$,
$\sim10^{3}$ $M_{\odot}$, and $\sim$ 3.3 $\times$ $10^{4}$ $M_{\odot}$
respectively. Similar amounts of gas are inferred to be in the
phase not well traced by the {\HI} or CO lines. Accumulation of more
gamma-ray data, particularly at high energies, and progress in ISM
studies, will reveal the CR and
matter distribution in greater detail.

\bigskip 
\begin{acknowledgments}
The Fermi LAT Collaboration acknowledges support from a number of
agencies and institutes for both development and the operation of the
LAT as well as scientific data analysis. These include NASA and DOE in
the United States, CEA/Irfu and IN2P3/CNRS in France, ASI and INFN in
Italy, MEXT, KEK, and JAXA in Japan, and the K. A. Wallenberg
Foundation, the Swedish Research Council and the National Space Board
in Sweden. Additional support from INAF in Italy and CNES in France
for science analysis during the operations phase is also gratefully
acknowledged.
\end{acknowledgments}

\bigskip 

\end{document}